\documentclass[a4paper, 12pt]{article}
\usepackage{a4}
\usepackage{amsfonts}
\usepackage{amssymb}
\usepackage{cite}
\usepackage{epsfig}
\usepackage{psfig}

\author{}

\newcommand{\ch}{{\rm ch}}
\newcommand{\be}{\begin{equation}}
\newcommand{\ee}{\end{equation}}
\newcommand{\bea}{\begin{eqnarray}}
\newcommand{\eea}{\end{eqnarray}}

\newcommand{\1}{{\bf 1}}

\newcommand{\Anti}{{\bf Anti}}
\newcommand{\Adj}{{\bf Adj}}
\newcommand{\N}{{\bf N}}

\newcommand{\ov}{\overline}
\def\IR{\relax{\rm I\kern-.18em R}}

\def\IP{\relax{\rm I\kern-.18em P}}
\def\inbar{\vrule height1.5ex width.4pt depth0pt}
\def\IC{\relax\,\hbox{$\inbar\kern-.3em{\rm C}$}}

\def\K3{{\bf K3}}

\def\ov{\overline}

\begin{document}

\title{
\begin{flushright} \vspace{-2cm}
{\small MPP-2005-70 \\
\vspace{-0.35cm}
hep-th/0507041} \end{flushright}
\vspace{4.0cm}
Supersymmetric (non-)Abelian Bundles in the\\
Type I and SO(32) Heterotic String
}
\vspace{1.0cm}
\author{\small Ralph~Blumenhagen, Gabriele Honecker and  Timo Weigand}
\date{}

\maketitle

\begin{center}
\emph{Max-Planck-Institut f\"ur Physik, F\"ohringer Ring 6, \\
  80805 M\"unchen, Germany } \\
\vspace{0.2cm}
\tt{blumenha,gabriele,weigand @mppmu.mpg.de}
\vspace{1.0cm}
\end{center}
\vspace{1.0cm}

\begin{abstract}
\noindent  
We discuss perturbative four-dimensional compactifications of both the
$SO(32)$ heterotic and the Type I string on
smooth Calabi-Yau manifolds endowed with 
general non-abelian and abelian bundles. 
We analyse  the generalized Green-Schwarz mechanism for multiple anomalous $U(1)$ factors
and derive the generically non-universal one-loop threshold
corrections to the gauge kinetic function 
as well as the one-loop corrected
Fayet-Iliopoulos terms.
The latter can be interpreted as 
a stringy one-loop correction to the Donaldson-Uhlenbeck-Yau
condition.
Applying S-duality, for the Type I string 
we  obtain  the perturbative $\Pi$-stability condition
for  non-abelian bundles on curved spaces.
Some simple examples are given, 
and we qualitatively discuss some generic phenomenological
aspects of this kind of string vacua. In particular, we point out
that in principle an intermediate  string scale scenario with  TeV scale large
extra dimensions might be possible
for the heterotic string.

\end{abstract}

\thispagestyle{empty}
\clearpage

\tableofcontents

\section{Introduction}

Over the past two decades, model building techniques have been developed 
in several corners of 
the M-theory moduli space. While early attempts focused on the heterotic 
$E_8 \times E_8$ string, 
the last ten years have seen considerable progress on the Type I side, 
both in the framework of 
intersecting D-branes (for a recent review see~\cite{Blumenhagen:2005mu} and 
references therein) 
and its T-dual formulation of magnetised D-branes.

In view  of the conjectured S-duality between the heterotic $SO(32)$ and 
Type I string theories, the concrete constructions studied so far in both frameworks 
exhibit a couple of features which at first sight appeared to be not quite compatible
\cite{Ibanez:1998qp,Lalak:1999bk}:
\begin{itemize}
\item{
The common perception seems to be that 
on the Type I side one generically encounters multiple anomalous
$U(1)$ gauge symmetries, whose anomalies are cancelled by a generalised Green-Schwarz
mechanism involving the K\"ahler-axion 
supermultiplets (also called internal axions in the remainder of this article) \cite{Ibanez:1998qp,Sagnotti:1992qw}.
For the heterotic string, by contrast, almost all previous models
show at most one anomalous $U(1)$, whose anomaly
is cancelled by the universal axion complexifying the
four-dimensional dilaton. 
This  includes heterotic orbifold, free-fermion and Gepner model
constructions.
}
\item{
For the known heterotic examples the Fayet-Iliopoulos (FI)
term vanishes identically  at string tree level, but often receives a non-vanishing contribution 
generated at one loop \cite{Dine:1987xk,Atick:1987gy}.
The latter  can only be cancelled
by giving vacuum expectation values (VEVs) to scalar fields charged
under the $U(1)$ in question. On the Type I side, the generic
string tree level
Fayet-Iliopoulos term is non-vanishing  but contains terms at 
leading
and next-to-leading order in $\alpha'$ \cite{Cremades:2002te}. 
Being also moduli dependent, it vanishes
for certain choices of the moduli without invoking 
any charged scalars.
}
\item{
In the heterotic constructions the gauge threshold corrections
for the non-abelian gauge factors are universal \cite{Nilles:1997vk}, 
whereas for the Type I string each stack of
D-branes comes with its own independent gauge coupling \cite{Blumenhagen:2000wh}.}
\item{The known heterotic string examples possess no adjoint scalars, which in turn are 
a common feature of the Type I string examples.}
\item{Last but not least, one has got used to the perception that on the heterotic side the string
 scale is always close to the Planck scale, whereas for D-brane models
the string scale is in principle a free parameter and the large 
Planck scale is an effect of large extra (transversal) dimensions.}
\end{itemize}

The point we would like to make is that these statements are not generically
true for all four-dimensional heterotic string compactifications, 
but are valid only under the restricted assumptions under which they
were derived. 
More concretely, in~\cite{Blumenhagen:2005ga} we argued that the
heterotic and Type I string vacua investigated so far were different
from the very beginning so that one could not expect
to get similar results.
On the Type I side one was considering various D-branes
equipped with abelian gauge bundles (magnetized branes),
whereas for the heterotic string one was mostly invoking
non-abelian gauge bundles on Calabi-Yau manifolds often
at their orbifold limits.
 
Therefore, in order to elucidate the possible difference
between Type I and heterotic models, one had
better make sure that one indeed compares S-dual constructions. 
This demonstrates the importance of studying heterotic string compactifications
on Calabi-Yau manifolds equipped also with abelian gauge bundles
\footnote{See \cite{Green:1984bx, Strominger:1986uh, Distler:1987ee, Aldazabal:1996du, Berglund:1998va,Andreas:2004ja} 
for earlier studies of models of this type.} 
as well as Type I strings with non-abelian vector bundles
on the world-volume of the D9-branes. In order
to avoid complications with orbifold singularities and
fixed point resolutions, it is much more transparent
to work on smooth Calabi-Yau spaces
and think about issues of orbifolds later.

In~\cite{Blumenhagen:2005ga}, building upon earlier work \cite{Lukas:1999nh}, 
it was shown for the $E_8 \times E_8$ heterotic 
string that multiple anomalous
$U(1)$-factors occur in the four dimensional compactifications if the compact 
six-dimensional manifold
is endowed with non-trivial line bundles. In this scenario, also the 
gauge kinetic function for the abelian observable gauge group factors
turns out
to be non-universal and there is a moduli dependent  non-vanishing tree-level
and one-loop contribution to the Fayet-Iliopoulos terms. 
This one-loop correction to the supersymmetry 
condition for the bundles allows for 
non-trivial models even on a Calabi-Yau with only one K\"ahler modulus, freezing a 
combination of the dilaton and the
K\"ahler modulus by supersymmetry.

The aim of this article is to continue along the lines of~\cite{Blumenhagen:2005ga} 
and to investigate four-dimensional 
Calabi-Yau compactifications of the heterotic $SO(32)$ string 
equipped with not only non-abelian but also abelian bundles. 
We elaborate on the general framework for constructing such models
in the perturbative, i.e. small string coupling and large
radii, regime. Special  emphasis will  be put  on clarifying the general
structure that arises in  the  sector of the abelian gauge symmetries. 
We will work out the Green-Schwarz mechanism and demonstrate how in general it invokes
both the universal and the internal axions.
We will also compute the tree-level and one-loop threshold corrected
gauge kinetic functions and find them to be non-universal. 
Furthermore, we will derive the perturbatively exact Fayet-Iliopoulos terms, which 
again contain a tree-level and a one-loop contribution. By S-duality
these are precisely related to the perturbative part
of the $\Pi$-stability condition \cite{Douglas:2000ah}
\bea
\label{pisus}\
 \xi\sim {\rm Im}\left(  \int_{\cal M} {\rm tr}_{U(N)}
\left[  e^{i\varphi}\, e^{J \, {\rm id}  +i {\cal F}}\,
         \sqrt{\hat A({\cal M})} \right]\right) =0
\eea
for non-abelian $U(N)$ 
bundles on manifolds with non-vanishing curvature.
To our knowledge the precise form of this expression has so far never been 
derived from first principles and we find it amusing
though not unexpected that it is already implicit in the
celebrated  Green-Schwarz anomaly terms \cite{Green:1984sg}.

The paper is organised as follows: We start in section~\ref{SecModel} by discussing the four-dimensional spectrum of heterotic $SO(32)$
compactifications for general $U(N)$ bundles and by computing the field 
theory anomalies.
In section~\ref{GSM}, the corresponding Green-Schwarz mechanism involving 
the dilaton-axion and K\"ahler-axion 
multiplets is discussed in very much detail. 
Section~\ref{SecGKF} contains the computation of the non-universal gauge kinetic function, and in section~\ref{SecFI},
the one-loop corrected supersymmetry condition is derived.
Section~\ref{SecSdual} is devoted to a discussion of S-duality to the Type I string with magnetised 
D-branes including an analysis of the appropriate stability conditions.
In section~\ref{SecEx}, we finally give two simple examples which are supposed
to serve merely as appetizers for the rich structure behind the new model building
perspectives. A more detailed study of realistic models remains to be endeavoured. In section \ref{SecPhen} we briefly address some phenomenological issues for this
class of heterotic string compactifications, restricting ourselves mostly to a qualitative level. In particular we point
out that for certain choices of the internal bundles it might
be possible to lower the string scale down to the
intermediate regime even for heterotic strings. 
Our conclusions are displayed in section~\ref{SecCon}.
Some technical details for the computations of sections \ref{SecModel} and \ref{GSM} are relegated to
appendices~\ref{AppEulerExp} and \ref{AppTrace}, respectively.

\section{Breaking of $SO(32)$ via unitary bundles}
\label{SecModel}

The aim of  this section is to define a general class of compactifications
of the $SO(32)$ heterotic string  involving 
direct sums of bundles with unitary structure groups.
This includes direct sums of abelian bundles, i.e.
line bundles, which have mainly been addressed
in the S-dual models of magnetised D-branes respectively
intersecting D-branes in the mirror symmetric setting.  
 
Concretely, we consider decompositions of $SO(32)$ into 
\bea
SO(2M) \times \prod_{x=1}^{K+L} U(N_x)
\eea
with \mbox{$M+\sum_{x=1}^{K+L} N_x=16$}.
One can realise such a breaking of the original $SO(32)$ by giving a background value to bundles on the internal 
manifold with structure group   
\mbox{$G=U(1)^{K} \times \prod_{m=K+1}^{K+L} U(N_m)$}, i.e. by turning on bundles of the form
\bea
\label{decomp1}
W = \bigoplus_{m=K+1}^{K+L} V_{m} \oplus \bigoplus_{i=1}^{K} L_i.
\eea
Note that in the following the index $i$ always runs
over the range $\{1,\ldots,K\}$, the index $m$ over
the range $\{K+1,\ldots,K+L\}$ and the index $x$
over the entire range $\{1,\ldots,K+L\}$.
The resulting observable non-abelian gauge group is 
\mbox{$H=SO(2M) \times \prod_{j=1}^K SU(N_j)$} and 
the anomaly free part of the $U(1)^{K+L}$ factors also remains in the low energy gauge group.

The second choice is to include the $SO(2M)$ factor in the bundle, i.e. to start with 
\bea
\label{decomp2}
W = V_0 \oplus  \bigoplus_{m=K+1}^{K+L} V_{m} \oplus \bigoplus_{i=1}^{K} L_i
\eea
with $c_1(V_0)=c_3(V_0)=0$
and the observable non-abelian part of the gauge group  $H=\prod_{j=1}^K SU(N_j)$. 

The general model building constraints at tree and stringy one-loop level for $E_8 \times E_8$ 
have been discussed at length in~\cite{Blumenhagen:2005ga}.  
Let us give the corresponding conditions for $SO(32)$ here:

\begin{itemize}
\item
In order to admit spinors, we need $c_1(W) \in H^2({\cal M}, 2{\mathbb Z})$.
\item
At string tree level, the field strength of the vector bundle has to satisfy the zero-slope limit of
the Hermitian Yang-Mills equations,
$F_{ab}=F_{\ov a\ov b}=0,$   $g^{a\ov b}\, F_{a\ov b}=0,$
constraining to $\mu$-stable, holomorphic vector bundles which satisfy the integrability condition        
\mbox{$\int_{\cal M}  J\wedge J \wedge c_1(V) = 0$}
for each constituent $V$  of the total bundle.
\item
The non-holomorphic Hermitian  Yang-Mills equation is modified at one loop. 
As we will compute in section~\ref{SecFI}, for a $U(N)$ bundle $V$ 
the perturbatively exact integrability condition reads
\bea
\label{DUYloop}
{1 \over 2}    \int_{\cal M} J\wedge J \wedge c_1(V)  - 
    g_s^2\, \ell_s^4 \, \int_{\cal M} 
\left( {\rm ch}_3 (V)+\frac{1}{24}\, c_1(V)\,  c_2(T)\right)=0,
\eea
where $g_s=e^{\phi_{10}}$ and $\ell_s=2\pi\sqrt{\alpha'}$.
An additional constraint arises from the requirement 
that the one-loop corrected $U(1)$ gauge couplings are real,
\bea
\label{Gauged}
    {N\over 3!}\,
 \int_{\cal M} J \wedge J \wedge J -
g_s^2\, \ell_s^4\,  \int_{\cal M} J \wedge 
\left( {\rm ch}_2 (V)+\frac{N}{24}\, c_2(T)\right) >0,
\eea
in the perturbative regime, i.e. for $g_s \ll 1$ and $r_i \gg 1$. 
Note that there will be additional stringy and $\alpha'$ non-perturbative corrections
to (\ref{DUYloop}) and (\ref{Gauged}).

\item
The Bianchi identity for the three-form 
$H=dB-{\alpha'\over 4}(\omega_{Y}-\omega_L)$ imposes the `tadpole condition'
\mbox{$dH={\alpha'\over 4}(  {\rm tr}(R^2)-{\rm tr}(F^2))$,}
where the traces are taken in the fundamental representation of $SO(1,9)$ and $SO(32)$, respectively.
For decompositions of the type~(\ref{decomp1}), this condition reads
\bea \label{TP}
0= c_2(T)+\sum_{i=1}^{K} N_i \; \ch_2(L_i) + \sum_{m=K+1}^{K+L} \ch_2(V_m)  
\eea
in cohomology.
\item
The chiral spectrum can be determined from the Euler characteristics of the various bundles ${\cal W}$ 
occurring in the decomposition of $SO(32)$,
\bea 
\label{RRH}
         \chi({\cal M},{\cal W})
         =\int_{\cal M}\left[ {\rm ch}_3({\cal W})+
           {1\over 12}\, c_2(T)\, c_1({\cal W}) \right].
\eea
\end{itemize}

In the remainder of this section, we present the generic chiral spectrum and compute the field theoretical four-dimensional anomalies.
In section~\ref{GSM}, we show that these anomalies are exactly cancelled by the dimensional reduction of the 
ten-dimensional kinetic and one-loop Green-Schwarz counter terms.

The above models can be studied for arbitrary $N_x$
\footnote{The decomposition of the adjoint of $SO(32)$ is the same as 
in the case where in Type IIA orientifold  theory all 
D6-branes first lie on top of the O6-planes
and then stacks of $N_x$ branes are rotated away from this position.}:
The adjoint representation of $SO(32)$ decomposes as follows 
\bea
{\bf 496} \longrightarrow
\left(\begin{array}{c} 
(\Anti_{SO(2M)})_0\\
\sum_{x=1}^{K+L} (\Adj_{SU(N_x)})_0 + (K+L ) \times (\1)_0\\
\sum_{x=1}^{K+L} (\Anti_{SU(N_x)})_{2(x)} + h.c.\\
\sum_{x < y}\left[ (\N_x,\N_y)_{1(x),1(y)} + (\N_x,\ov{\N}_y)_{1(x),-1(y)} + h.c. \right]\\
\sum_{x=1}^{K+L} (2M, \N_x)_{1(x)} + h.c.
\end{array}\right).   
\eea
The massless spectrum therefore takes the form given in Table~\ref{Tchiral1} for bundles of type~(\ref{decomp1}).
For bundles of the form~(\ref{decomp2}), the states transforming under $SO(2M)$ are computed from Table~\ref{Tchiral2}. 
\begin{table}[htb]
\renewcommand{\arraystretch}{1.5}
\begin{center}
\begin{tabular}{|c||c|}
\hline
\hline
reps. & $H=\prod_{j=1}^K U(N_j) \times U(1)^{L} \times SO(2M)$   \\
\hline \hline
$(\Adj_{U(N_j)})_{0(j)}$ & $H^*({\cal M},{\cal O})$  \\
$(\1)_{0(m)}$ &  $H^*({\cal M},V_m \otimes V_m^{\ast} )$ \\
\hline
$(\Anti_{SU(N_j)})_{2(j)}$ & $H^*({\cal M},L_j^2)$  \\
$(\1)_{2(m)}$ & $H^*({\cal M},\bigwedge^2 V_m)$ \\
\hline
$(\N_i,\N_j)_{1(i),1(j)}$ &  $H^*({\cal M},L_i \otimes L_j)$  \\
$(\N_i,\ov{\N}_j)_{1(i),-1(j)} $ &  $H^*({\cal M},L_i \otimes L_j^{-1})$ \\
$(\N_i)_{1(i),1(m)}$ & $H^*({\cal M}, V_m \otimes L_i)$ \\
$(\N_i)_{1(i),-1(m)} $ &  $H^*({\cal M},V_m^{\ast} \otimes L_i)$  \\
$(\1)_{1(m),1(n)}$ & $H^*({\cal M}, V_m \otimes V_n)$ \\
$(\1)_{1(m),-1(n)} $ &  $H^*({\cal M}, V_m \otimes V_n^{\ast})$ \\
\hline
$(\Adj_{SO(2M)})$ & $H^*({\cal M},{\cal O})$ \\

$(2M, \N_j)_{1(j)}$ & $H^*({\cal M}, L_j)$\\
$(2M )_{1(m)}$ & $H^*({\cal M},V_m )$ \\ 
\hline
\end{tabular}
\caption{\small Massless spectrum. $M \equiv 16-\sum_{x=1}^{K+L}N_x$. The indices run over the range $i,j \in \{1,\ldots,K\}$ and
$m,n \in \{K+1,\ldots,K+L\}$. The structure group is taken to be $G=U(1)^{K} \times \prod_{m=K+1}^{K+L} U(N_m)$.  }
\label{Tchiral1}
\end{center}
\end{table}
\begin{table}[htb]
\renewcommand{\arraystretch}{1.5}
\begin{center}
\begin{tabular}{|c||c|}
\hline
\hline
reps.   &  $H=\prod_{j=1}^K U(N_j) \times U(1)^{L}$  \\
\hline \hline
$(\1)$ & $H^*({\cal M},\bigwedge^2 V_0  )$ \\
 $(\N_j)_{1(j)}$ &  $H^*({\cal M}, V_0 \otimes L_j)$ \\
$(\1)_{1(m)}$ &  $H^*({\cal M},V_0 \otimes V_m )$\\ 
\hline
\end{tabular}
\caption{\small Massless spectrum for $G=U(1)^{K} \times \prod_{m=K+1}^{K+L} U(N_m)\times SO(2M)$. These three lines replace  
the last three lines in Table~\ref{Tchiral1}, all other cohomology classes are identical in both cases.}
\label{Tchiral2}
\end{center}
\end{table}

Note that this is the same massless spectrum as for the perturbative
Type I string on a smooth Calabi-Yau space with B-type D9-branes. 
In particular, turning on abelian bundles for  the $SO(32)$ heterotic string
also leads to $H^1({\cal M}, {\cal O})$ 
massless chiral multiplets  transforming in the {\it adjoint}
representation of a $U(N_j)$ observable gauge factor. 
In many cases there do not exist any non-trivial homological one-cycles but
on the torus one has for instance $H^1(T^6, {\cal O})=3$.
These complex adjoint scalars correspond to the continuous Wilson lines on ${\cal M}$.
Analogously, turning on non-abelian bundles on the
Type I D9-branes gives rise to 
$H^1({\cal M},(V_m \otimes V_m^{\ast}))$ moduli corresponding to the
deformations of the $U(N_m)$ bundle.

Now let us discuss the resulting anomalies. 
We will first consider  the case with structure group $G=U(1)^{K} \times \prod_{m=K+1}^{K+L} U(N_m)$
and the resulting observable gauge group 
\mbox{$H=SO(2M) \times \prod_{j=1}^K SU(N_j) \times U(1)^{K+L} $}.
The anomalies are computed from the net number of chiral multiplets of the diverse representations ${\cal W}$
using~(\ref{RRH}).
For the cubic non-abelian anomalies we obtain from
\bea
{\cal A}_{SU(N_i)^3} &\sim&
(N_i-4)\chi(L_i^2) + \sum_{j \neq i} N_j \left(\chi(L_i \otimes L_j) + \chi(L_i \otimes L_j^{-1})   \right)  
\nonumber \\
&&+\sum_m \left(\chi(V_m \otimes L_i) +\chi(V_m^{\ast} \otimes L_i)      \right)
+ 2M \chi (L_i),
\eea
the expression in terms of Chern characters,
\bea
{\cal A}_{SU(N_i)^3} &\sim& 2 \int_{\cal M} c_1(L_i) \times {\rm Tad},
\eea
with the tadpole condition  
\bea \label{Tad1}
{\rm Tad} = c_2(T)+\sum_{j=1}^{K} N_j \; \ch_2(L_j) + \sum_{m=K+1}^{K+L} \ch_2(V_m)=0  
\eea
in cohomology.
Thus in contrast to the $E_8 \times E_8$ examples discussed in~\cite{Blumenhagen:2005ga}, the cubic
non-abelian anomalies vanish only upon tadpole cancellation\cite{Witten:1984dg}.

The explicit expressions for all mixed and cubic abelian anomalies in terms of 
Euler characteristics are given in appendix~\ref{AppEulerExp}. Here we only state the result in 
terms of the various Chern characters up to tadpole cancellation for the $U(1)_i$ factors with $i\in\{1,\ldots,K\}$
\bea
{\cal A}_{U(1)_i-SU(N_j)^2} &\sim& \int_{\cal M}  N_i \;  c_1(L_i) \left(\frac{1}{6}\,
c_2(T) +2\, \ch_2(L_j) \right)
+ 2 \int_{\cal M}  N_i \; \ch_3(L_i),
 \nonumber \\
{\cal A}_{U(1)_i-G^2_{\mu\nu}} &\sim& \frac{1}{2}\int_{\cal M} 
 N_i \;  c_1(L_i)\, c_2(T) + 24\int_{\cal M} N_i \; \ch_3(L_i),
 \nonumber \\
{\cal A}_{U(1)_i-U(1)_j^2} &\sim&  N_j\int_{\cal M}
 N_i \;c_1(L_i)\left(\frac{1}{6}\, c_2(T) +2\ch_2(L_j) \right)
 +  2\, N_j\int_{\cal M} N_i \;\ch_3(L_i),
 \nonumber \\
{\cal A}_{U(1)_i -SO(2M)^2} &\sim &  \frac{1}{12}\int_{\cal M}
 N_i \; c_1(L_i)\, c_2(T)+  \int_{\cal M}  N_i \; \ch_3(L_i).
\label{EqAnom}
\eea
The corresponding expressions for the $U(1)_m$ factors with $m\in\{K+1,\ldots,K+L\}$ are obtained by replacing
everywhere $N_i \;\ch_k (L_i)$ by $\ch_k (V_m)$.  This reflects the fact that in the latter case, we 
have internal $U(N_m)$ bundles, while for $i\in\{1,\ldots,K\}$, the $SU(N_i)$ factor is external and 
appears only as an overall factor $N_i$ in all Chern characters.

The field theory anomalies in the
 case where $SO(2M)$ is part of the bundle have the same form (\ref{EqAnom})
after the modified tadpole condition, 
\bea
\widehat{\rm Tad} = c_2(T)+\sum_{i=1}^{K} N_i \ch_2(L_i) + \sum_{m=K+1}^{K+L} \ch_2(V_m)   - \frac{1}{2} c_2(V_0)=0,  
\eea
is taken into account.
The mixed anomalies ${\cal A}_{U(1)_x-SO(2M)^2}$ are of course absent in this case.


\section{Green Schwarz mechanism for the heterotic $SO(32)$ string in 4D}
\label{GSM}

The anomalies encountered in the four-dimensional effective field theory have to be cancelled by a 
generalised 
Green-Schwarz (GS) mechanism for consistency of the models. As we show here, the couplings of the 
$h_{11}+1$ 
four-dimensional axions to the gauge fields and gravity arising from the ten-dimensional 
kinetic and one-loop counter term have precisely the correct form. In addition,  
we will explicitly derive the mass terms for the various axionic fields in order to determine which 
linear combinations of $U(1)$s remain massless in the low-energy effective field theory.

Let us first discuss the couplings which arise from the dimensional reduction of the 
ten-dimensional Green-Schwarz one-loop counter term \cite{Green:1984sg}
($\ell_s \equiv 2\pi \sqrt{\alpha'}$)  
\bea
S_{GS} &=& \frac{1}{48(2\pi)^3 \ell_s^2} \int B^{(2)} \wedge X_8
\eea
with the eight-form given by
\bea   
  X_8={1\over 24} {\rm Tr} F^4 -{1\over 7200} \left( {\rm Tr} F^2\right)^2 
      -{1\over 240} \left( {\rm Tr} F^2\right) \left( {\rm tr} R^2\right)+
       {1\over 8}{\rm tr} R^4 +{1\over 32} \left( {\rm tr} R^2\right)^2 
\eea
 for both the $SO(32)$ and $E_8 \times E_8$ heterotic theories where ${\rm Tr}$ denotes the trace 
in the 
adjoint representation and ${\rm tr}$ the 
one in the fundamental. Although the eight-form is the same for $SO(32)$ and $E_8 \times E_8$, 
the resulting four-dimensional effective couplings turn out to differ since in the latter case
there is no independent fourth order Casimir.

In the $SO(32)$ case, dimensional reduction of the GS counter term to four dimensions gives
\bea
S_{GS}&=& \frac{1}{(2\pi)^3 \ell_s^2}  
\int B  \wedge  \frac{1}{288} {\rm Tr} (F \ov F^3)  \label{GSmass1} \\
&-& \frac{1}{(2\pi)^3 \ell_s^2}\int B  \wedge  \frac{1}{5760} {\rm Tr} (F {\ov F})  
\wedge \left( \frac{1}{15}  {\rm Tr} {\ov F}^2 
+ {\rm tr} \ov R^2 \right)  \label{GSmass2}\\
&+& \frac{1}{(2\pi)^3 \ell_s^2}\int B \wedge \left( \frac{1}{192} {\rm Tr}( F^2 {\ov F}^2 ) - 
\frac{1}{86400}  [{\rm Tr} (F {\ov F})]^2 \right) \label{GSvF1} \\ 
&-& \frac{1}{(2\pi)^3 \ell_s^2}\int  B \wedge \frac{1}{11520}  {\rm Tr} (F^2) \wedge  
\left( \frac{1}{15} {\rm Tr} { \ov F}^2 +  {\rm tr} {\ov R}^2 \right)
\label{GSvF2}  \\
&+&\frac{1}{(2\pi)^3 \ell_s^2} \int  B \wedge  \frac{1}{768} {\rm tr} { R}^2 
\wedge \left( {\rm tr} {\ov R}^2 - \frac{1}{15}  {\rm Tr} { \ov F}^2 \right),
\label{GSR}
\eea
where we denote by $\ov F$ gauge fields taking values along the compact directions and $F$ along the 
four-dimensional
non-compact ones. The two-form $B$ has both legs chosen along the non-compact directions in the 
first two lines and along the  compact directions
in the remaining three terms.
The expressions (\ref{GSmass1}), (\ref{GSmass2}) are mass terms for the $U(1)$ gauge factors. 
(\ref{GSvF1}) and (\ref{GSvF2}) lead to
vertex couplings of the axions with  two gauge fields and 
finally  the expression  (\ref{GSR}) gives rise  to   vertex couplings of the axions and 
two gravitons.

As in~\cite{Blumenhagen:2005ga}, there are additional  mass terms and vertex couplings arising from the kinetic term in the 
ten-dimensional effective action 
\bea
\label{kinH}
         S_{kin}=-{1\over 4\, \kappa_{10}^2}\, \int  e^{-2\phi_{10}}\,
       H\wedge \star_{10}\, H,
\eea
where $\kappa^2_{10}={1\over 2}(2\pi)^7\, (\alpha')^4$ and
the heterotic 3-form field  
strength  $H=dB^{(2)}-{\alpha'\over 4}(\omega_{Y}-\omega_{L})$ 
involves the 
gauge and gravitational Chern-Simons terms.
In terms of  the six-form $B^{(6)}$ dual to $B^{(2)}$, $\star_{10}\, dB^{(2)}=e^{2\phi_{10}}\,dB^{(6)}$, (\ref{kinH}) contains
\bea
\label{kin2}
S_{kin} &=& \frac{1}{8\pi \ell_s^6} \int \left( {\rm tr} F^2 - {\rm tr} R^2 \right)\wedge B^{(6)}.  
\eea

The discussion of the GS mechanism upon compactification necessitates the extraction of the various 
vertex couplings and mass terms from the above contributions to the four-dimensional Lagrangian. 
It is convenient to introduce  a basis $\omega_k$, $(k= 1, \ldots, h_{11})$  of 
$H^2({\cal M},{\mathbb Z})$ 
together with the Hodge dual four-forms $\widehat{\omega}_{k}$ normalised such that 
$\int_{\cal M}  \omega_k  \wedge \widehat{\omega}_{k'} = \delta_{k, k'}$. In terms of these, 
the expansion of the various fields is
\bea
    B^{(2)}&=& b^{(2)}_0+\ell_s^2\, 
          \sum_{k=1}^{h_{11}}   b^{(0)}_k\, \omega_k ,\quad
    {\rm tr} \ov {F}^2
= (2\pi)^2\, \sum_{k=1}^{h_{11}} 
({\rm tr} \ov {F}^2)_k
\, \widehat{\omega}_{k}, \nonumber \\
   \ov f^m&=&2\pi\, \sum_{k=1}^{h_{11}}   \ov f^m_k\, \omega_k ,\quad\phantom{aaaa}
     {\rm tr} \ov R^2=(2\pi)^2\, \sum_{k=1}^{h_{11}} 
      \bigl( {\rm tr} \ov R^2\bigr)_k\, \widehat{\omega}_{k}, \nonumber\\
 \label{EexpB6}
      B^{(6)}&=&\ell_s^6\, b^{(0)}_0\,  {\rm vol}_6 +
      \ell_s^4\,
           \sum_{k=1}^{h_{11}}  b^{(2)}_k\, \widehat\omega_k.
\eea
Here ${\rm vol}_6$ represents the volume form of the internal Calabi-Yau manifold normalised such 
that $\int_{\cal M} {\rm vol}_6 = 1$.

The traces occurring in the kinetic and counter terms have to be evaluated for a  concrete choice of 
internal bundle. 
The results  for the internal bundle as given in (\ref{decomp1}) are listed in Appendix B, where we 
also summarise our notation used in the remainder of this paper. 
With these results at hand, it is a simple task to collect the explicit mass and GS terms. 

From (\ref{GSmass1}), (\ref{GSmass2}) we find that the four-dimensional two-form field $ b_0^{(2)}$ is 
rendered massive by the coupling to the abelian gauge fields given by
\bea
\label{mass2}
S^0_{mass}
&=&  \sum_{x=1}^{K+L} \frac{ 1 }{6 (2 \pi)^5 \alpha'} \int_{{\mathbb R}_{1,3}} b_0^{(2)} \wedge  
f_x \int_{\cal M} \Bigl( {\rm tr} {\ov {F}}^3_x - \frac{1}{16} 
 {\rm tr}{\ov {F}}_x \wedge  {\rm tr}{\ov R}^2  \Bigr), 
\eea
where we denote by ${\rm tr} ({\ov {F}}_x^n) $ the trace in the fundamental representation of 
the  $U(N_x)$-bundle from which the abelian factor $ f_x$ descends. Again, in (\ref{mass2}) and in all 
expressions which follow  it is understood that the non-abelian part of ${\ov {F}}_x $  is 
present only for $x \in \{K+1, \ldots, K+L\}$. Traces over the fundamental representation of
an $SU(N_i)$ factor are denoted by ${\rm tr}_i (\widehat{F}^n)$ for distinction.

In addition, (\ref{kin2}) gives rise to mass terms for the internal two-forms  $b_k^{(2)}$,
\bea
\label{kin3} 
S_{mass} = \sum_{x=1}^{K+L}\sum_{k=1}^{h_{11}} \frac{1}{ (2 \pi)^2 \alpha'}  
\int_{{\mathbb R}_{1,3}}(  b_k^{(2)} \wedge  f_x)  ({\rm tr}{\ov {F}}_x)_k.
\eea

As for the gauge  anomalies, the GS counter terms (\ref{GSvF1}) and (\ref{GSvF2}) provide the 
anomalous couplings of the axions to the 
external gauge fields,
\bea
S_{GS}&=& \frac{1}{ 2 \pi} \int_{{\mathbb R}_{1,3}} \sum_{k=1}^{h_{11}}\sum_{i=1}^K  b_k^{(0)}\,  {\rm
  tr}_i \widehat{F}^2\,  \Bigl( \frac{1}{4} \,  ({\ov f}_i^2)_k - \frac{1}{192} ({\rm tr} {\ov R}^2)_k \Bigr)
\label{GSFi2}\\
&& 
-\frac{1}{384}\frac{1}{ 2 \pi} \int_{{\mathbb R}_{1,3}} \sum_{k=1}^{h_{11}}
 b_k^{(0)}\,   {\rm tr}_{SO(2M)} F^2\,  ({\rm tr} {\ov R}^2)_k
\label{GSFSO2}\\
&&
+\frac{1}{ 2 \pi} \int_{{\mathbb R}_{1,3}} \sum_{k=1}^{h_{11}}
\sum_{x=1}^{K+L} b_k^{(0)}\,  f_x^2\, 
\Bigl(\frac{1}{4} \,  ({\rm tr}{\ov {F}^2}_x)_k -\frac{N_x}{192}({\rm tr} {\ov R}^2)_k \Bigr).
\label{GSfx2}
\eea
Combining them  with (\ref{kin3}) and  taking into account the analogous vertex couplings
\bea
\label{kinvF}
S_{GS}^0 = \frac{1}{8 \pi} \int_{{\mathbb R}_{1,3}} b_0^{(0)}\,  \Bigl(2
\sum_{i=1}^K{\rm tr}_i \widehat{F}^2 +{\rm tr}_{SO(2M)} F^2+2\sum_{x=1}^{K+L} N_x\, f_x^2  \Bigr)
\eea
from (\ref{kin2}) together with its counterpart (\ref{mass2}) leads to an expression for the 
respective anomaly six-form. For the mixed $U(1)_x - SU(N_i)$ anomaly, for instance, we find
\bea
{\cal A}_{U(1)_x -SU(N_i)^2  } &\sim& \frac{1}{12  (2 \pi)^5 \alpha'} \,  f_x \wedge  
{\rm tr}_i \widehat{F}^2 \nonumber \\
 && \int_{\cal M} \Bigl( {\rm tr}  {\ov {F}}^3_x + 3 \; {\rm tr} {\ov {F}}_x \wedge 
{\ov f}_i^2  - \frac{1}{8} {\rm tr}{\ov {F}}_x 
\wedge {\rm tr} {\ov R}^2 \Bigr),   
\eea
which is just tailor-made to cancel the mixed  $U(1)_x - SU(N_i)^2$ anomaly. 
The cancellation pattern for the remaining abelian-non-abelian, cubic abelian and  mixed gravitational 
anomalies follows the same lines. In the latter case the internal axion-graviton couplings 
follow from (\ref{GSR}) and the external one from~(\ref{kin2}).
Let us just list the resulting anomaly six-forms
\bea
{\cal A}_{U(1)_x - SO^2}\!\!\!&\sim&\!\!\! \frac{1}{24 (2 \pi)^5 \alpha'} \,  f_x \wedge  
{\rm tr}_{SO(2M)} F^2  \nonumber 
 \int_{\cal M} \Bigl( {\rm tr}  {\ov {F}}^3_x  - \frac{1}{8} {\rm tr}{\ov {F}}_x 
\wedge{\rm tr} {\ov R}^2 \Bigr), \\
{\cal A}_{U(1)_x - G_{\mu \nu}^2}\!\!\! &\sim&\!\!\! - \frac{1}{24 (2 \pi)^5 \alpha'} f_x 
\wedge  {\rm tr} R^2  
 \int_{\cal M} \Bigl( {\rm tr}  {\ov {F}}^3_x  - \frac{1}{16} {\rm tr}{\ov {F}}_x 
\wedge{\rm tr} {\ov R}^2 \Bigr), \\
{\cal A}_{U(1)_x - U(1)_y^2} \!\!\!&\sim&\!\!\! \frac{1}{12 (2 \pi)^5 \alpha'} f_x \wedge f_y^2  
\nonumber 
 \int_{\cal M}  \Bigl( N_y\, ({\rm tr}  {\ov {F}}^3_x  - \frac{1}{8} {\rm tr}{\ov {F}}_x 
\wedge {\rm tr}{\ov R}^2) +  {\rm tr}{\ov {F}}_x \wedge{\rm tr}{\ov {F}}_y ^2        \Bigr)
\eea
and point out that they are in perfect agreement with the field theoretic anomalies given in the previous section.
As usual,  the anomalous $U(1)$s are rendered massive and therefore remain in the low-energy domain 
as perturbative global symmetries. The situation parallels that in Type I\cite{Ibanez:2001nd} and 
heterotic $E_8 \times E_8$-theory \cite{Blumenhagen:2005ga}, where the number of massive abelian 
factors is at least as large as that of the anomalous ones and in general given by the rank of the 
mass matrix
\bea
\label{massmatrix} 
    {\rm M}_{xk}=\cases{    \frac{1}{ (2 \pi)^2 \alpha'}   ({\rm tr}{\ov {F}}_x)_k    
& for $k\in\{1,\ldots,h_{11}\}$ \cr
                    \frac{ 1 }{6 (2 \pi)^5 \alpha'} \int_{\cal M} \Bigl( {\rm tr} {\ov {F}}^3_x -
 \frac{1}{16} 
 {\rm tr}{\ov {F}}_x \wedge  {\rm tr}{\ov R}^2  \Bigr)   &
                   for $k=0$. \cr} 
\eea

\section{Non-universal gauge kinetic functions}
\label{SecGKF}

Let us now derive the gauge kinetic functions 
\cite{Derendinger:1985cv,Ibanez:1986xy,Nilles:1997vk,Stieberger:1998yi}.
The holomorphic gauge kinetic function ${\rm f}_a$ appears in the four-dimensional effective field 
theory as 
\bea
       {\cal L}_{YM}={1\over 4}\, {\rm Re} ({\rm f}_a)\, F_a\wedge\star F_a + {1 \over 4}\,
         {\rm Im} ({\rm f}_a)\, F_a\wedge  F_a. 
\eea
With the definition of the complexified dilaton and  K\"ahler moduli
\bea
    S={1\over 2\pi}\left[ e^{-2\phi_{10}} {{\rm Vol}({\cal M}) \over
             \ell_s^6 } + {i}\, b^{(0)}_0 \right],
\qquad\qquad
    T_k={1\over 2\pi}\left[ -\alpha_k + {i} b^{(0)}_k \right],
\eea
the gauge kinetic functions can be read off from their imaginary parts in~(\ref{GSFi2})-(\ref{kinvF}) 
to be
\bea
\label{ggg}
{\rm f}_{SU(N_i)} &=& 2S +\sum_{k=1}^{h_{11}} T_k\, \Bigl((\ov f_i^2)_k-\frac{1}{48}({\rm tr} \ov R^2)_k \Bigr), 
\label{Gaugecoupling}\\
{\rm f}_{SO(2M)} &=& S -\frac{1}{96}\sum_{k=1}^{h_{11}} T_k\, ({\rm tr} \ov R^2)_k,  \nonumber\\
{\rm f}_{x} &=& 2N_x S +\sum_{k=1}^{h_{11}} T_k\,  \Bigl(({\rm tr} \ov{F}^2_x)_k -\frac{N_x}{48}({\rm tr} 
\ov R^2)_k \Bigr). \nonumber
\eea

Note that the real parts of the gauge kinetic function
are  positive definite by definition. Therefore, requiring positivity 
of the expressions (\ref{ggg}) in the perturbative regime, $g_s\ll 1$ and $r_i\gg 1$,  
imposes  extra conditions on the allowed bundles. 
Away from the small coupling and large radii limit one expects both world-sheet and
stringy instanton corrections to the gauge kinetic functions \cite{Nilles:1997vk}.

The real parts of the K\"ahler moduli are defined by the expansion of the K\"ahler form in terms of the chosen
basis of two-cycles, $J=\ell_s^2 \sum_{i=1}^{h_{11}} \alpha_i \omega_i$, and the compact volume is computed from
\bea
      {\rm Vol}({\cal M})={1\over 6}\int_{\cal M} J\wedge J\wedge J=
     {  \ell_s^6 \over 6}\, \sum_{i,j,k} d_{ijk}\,  \alpha_i\, \alpha_j\, \alpha_k,
\eea 
where $d_{ijk} = \int_{\cal M} \omega_i \wedge \omega_j \wedge \omega_k$ are the triple intersection numbers of the basis of two-forms.
The explicit computation of the real parts is more involved, however, the coupling to the dilaton can be directly seen from the dimensional
reduction of the kinetic term of the gauge field,
\bea
     S^{(10)}_{YM}={1\over 2\kappa_{10}^2} \int e^{-2\phi_{10}} \, {\alpha'\over 4}
             {\rm tr}( F\wedge\star_{10} F).
\eea
For $i\in\{1,\ldots,K\}$, the gauge kinetic function for the abelian factor of the $U(N_i)$ group is proportional to the
non-abelian part, \mbox{${\rm f}_{U(1)_i}=N_i\,  {\rm f}_{SU(N_i)}$}.

Note that in contrast to the $E_8 \times E_8$ examples
in~\cite{Blumenhagen:2005ga}, no off-diagonal couplings among abelian factors
occur.
Even more strikingly, the tree-level and one-loop corrected
non-abelian and abelian gauge couplings of an observable $SU(N_i)$ and $U(1)_x$ gauge factor only depend
on the internal gauge flux in the same $U(N_i)$ and 
$U(N_x)$ gauge group respectively.  
Since we used the same decomposition of $SO(32)$
that naturally appears for intersecting D-branes,
S-duality tells us that after all  this result is not surprising.
There, each stack of D-branes comes with its own gauge coupling
determined by the size of the three-cycle the D6-branes
are wrapping around.

\section{Fayet-Iliopoulos terms and supersymmetry conditions}
\label{SecFI}

The appearance of anomalous $U(1)$ symmetries indicates the potential 
creation of Fayet-Iliopoulos (FI) terms in the four-dimensional effective action
\cite{Dine:1986zy,Dine:1987bq,Atick:1987gy}, 
which  
have to vanish for supersymmetry to be preserved. Note that in the following
we ignore the option to cancel a non-vanishing FI-term by giving 
VEVs to charged scalars. In fact, we strongly believe that in more realistic scenarios which include fluxes, the induced mass terms for these fields will fix their VEVs at zero independently as is the case in the extensively studied Type IIB models \cite{GarciadelMoral:2005js}.
From the general analysis of four-dimensional ${\cal N}=1$ supergravity it is well-known that the 
coefficients $\xi_x$ of the FI-terms can be derived from 
the K\"ahler potential $\cal K$ via the relation
\bea
\label{FIterms}
   {\xi_x\over g_x^2}=   {\partial {\cal K}\over \partial V_x}
\biggr\vert_{V=0}.
\eea
The gauge invariant K\"ahler potential relevant for our type of construction reads
\bea
{\cal K} &=&{M^2_{pl}\over 8\pi} \Biggl[   
       -\ln\Biggl(S+S^*-\sum_x Q^x_0\, V_x\Biggr)-\ln\Biggl(-\sum_{i,j,k=1}^{h_{11}}
{d_{ijk}\over 6} 
\biggl(  T_i+T_i^*-\sum_x Q^x_i\, V_x\biggr) \nonumber \\
&& \phantom{aaaaaaa} \biggl(  T_j+T_j^*-\sum_x Q^x_j\, V_x\biggr)
\biggl(  T_k+T_k^*-\sum_x Q^x_k\, V_x\biggr) \Biggr) \Biggr]
\eea
in the notation of \cite{Blumenhagen:2005ga}. The charges $ Q^x_k$ are
defined via 
\bea
                 S_{mass}=\sum_{x=1}^{K+L}  \sum_{k=0}^{h_{11}} 
            {Q^x_k\over 2\pi\alpha'}
                \int_{\IR_{1,3}} f_x \wedge b^{(2)}_k 
\eea
and  can easily be read off from the mass terms (\ref{mass2}) and  (\ref{kin3}).

The FI terms are seen to be proportional to the specific combination 
\bea
 {\xi_x\over g_x^2} = -\frac{e^{2 \phi_{10}} M_{pl}^2 \,\ell_s^6}{4  {\rm Vol}({\cal M})}
\Bigl(\frac{1}{4} e^{-2 \phi_{10}} \sum_{i,j,k=1}^{h_{11}} d_{ijk} Q^x_i \alpha_j \alpha_k - \frac{1}{2} Q^x_0 \Bigr).
\eea
Inserting the concrete expressions for the charges immediately leads to the conclusion that the FI terms vanish if and only if
\bea
\label{EDUY1}
e^{-2 \phi_{10}} \frac{1}{2} \int_{\cal M} J \wedge J \wedge { \rm tr}  {\ov {F}}_x  - \frac{( 2\pi \alpha')^2}{3 !}
\int_{\cal M} \Bigl( { \rm tr}  {\ov {F}}_x^3 - \frac{1}{16}  {\rm tr}  {\ov {F}}_x \wedge {\rm tr}{\ov R}^2 \Bigr)= 0
\eea
for each external $U(1)_x$ factor separately. 
Note again  that, as expected from the intersecting D-brane picture,
the FI-term for $U(1)_x$ only depends on the internal vector bundle
with field strength  $\ov{F}_x$.
In contrast to the $E_8\times E_8$ heterotic string,
there appears the  cubic term ${ \rm tr}  {\ov {F}}_x^3$
in the one-loop correction to the FI-term. This can be traced
back to the fact that in contrast to $E_8$ the group $SO(32)$ 
has an independent
fourth order Casimir operator. 
It implies the well-known result that for the $SO(32)$ heterotic
string a bundle with structure group $SU(N)$ generates
a non-vanishing one-loop FI-term \cite{Dine:1987xk}. 
Again, away from the small string coupling and large radii limit one expects
additional non-perturbative world-sheet and string instanton 
contributions to (\ref{EDUY1}).

\section{S-duality to the Type I string}
\label{SecSdual}

So far we have derived all equations for the $SO(32)$ heterotic
string. We will now apply Heterotic-Type I S-duality to
these equations and compare with known results on the Type I side. 

\subsection{The gauge couplings for Type I}

First let us write the expression for the gauge couplings
in a way which is more suitable for the S-duality transformation.
The real part of the holomorphic gauge kinetic function ${\rm f}_x$
can be cast into the form
\bea
{\rm Re}({\rm f}^H_{x}) = {1\over \pi\ell_s^6}\left[ {N_x\over 3!}\,  g_s^{-2}
 \int_{\cal M} J \wedge J \wedge J -
{(2\pi\alpha')^2\over 2} \int_{\cal M} J \wedge 
       \left( {\rm tr} \ov {F}_x^2  - {N_x \over 48}  {\rm tr}{\ov R}^2 \right)
\right].
\eea
Applying  the heterotic-Type I string duality relations 
\cite{Polchinski:1995df}
\bea
\label{ESduality}
         g_s^I&=& (g_s^H)^{-1}, \nonumber \\
            J^I&=&(g_s^H)^{-1} J^H
\eea   
with $g_s=e^{\phi_{10}}$  leads to 
\bea
\label{gal}
{\rm Re}({\rm f}^I_{x}) = {1\over \pi\ell_s^6 g_s}\left[ {N_x\over 3!}\, 
 \int_{\cal M} J \wedge J \wedge J -
{(2\pi\alpha')^2\over 2} \int_{\cal M} J \wedge 
       \left( {\rm tr} \ov {F}_x^2  - {N_x \over 48}  {\rm tr}{\ov R}^2 \right)
\right]
\eea
on the Type I side. Note that the second term has  now become a  perturbative $\alpha'$-correction to the tree-level gauge coupling.

\subsection{The non-abelian MMMS condition}

The same S-duality relations 
(\ref{ESduality}) applied to the FI-terms (\ref{EDUY1}) yield 
\bea
\label{fial}
\frac{1}{2} \int_{\cal M} J \wedge J \wedge { \rm tr}  {\ov {F}}_x  - \frac{( 2\pi \alpha')^2}{3 !} 
\int_{\cal M} \left( { \rm tr}  {\ov {F}}_x^3 - \frac{1}{16}  {\rm tr}  {\ov {F}}_x \wedge {\rm tr}{\ov R}^2\right) = 0
\eea
on the Type I side, where the second term is again a perturbative $\alpha'$-correction. 
We can combine the gauge kinetic function and the FI-term into a single 
complex quantity, the central charge
\bea
\label{cc}
{\cal Z} =  \int_{\cal M} {\rm tr}_{U(N)}
\left[\left( e^{J \, {\rm id}  +i {\cal F}}\,
         \sqrt{\hat A({\cal M})}\right) \right], \\ \nonumber
\eea
defined in terms of ${\cal F}=2\pi\alpha' \ov {F}$ and the A-roof genus 
$\hat A({\cal M})=1+\frac{1}{48}{1\over (2\pi)^2}{\rm tr}{\cal R}^2 +\cdots$ with ${\cal R}=\ell_s^2\, \ov R$. 
The gauge coupling  and the FI-term are seen to be proportional to 
the real and imaginary part, respectively, of ${\cal Z}$. 

In the case of  abelian D9-branes in Type IIB
we know 
that one can introduce an additional phase parameterising which
${\cal N}=1$  supersymmetry of the underlying  ${\cal N}=2$  supersymmetry 
is preserved by the brane.
Therefore, the general Type IIB supersymmetry condition 
is 
\bea
\label{SUSY}\
 {\rm Im}\left(  \int_{\cal M} {\rm tr}_{U(N)}
\left[  e^{i\varphi}\, e^{J \, {\rm id}  +i {\cal F}}\,
         \sqrt{\hat A({\cal M})} \right]\right) =0.
\eea
As usual in Type IIB theory coupled to a brane, we have now defined ${\cal F}=2\pi\alpha' \ov {F}+ B\, {\rm id}$,
thus taking  into account the fact that for open strings
only this combination  is a gauge invariant quantity.

Note that
$(\ref{cc})$
 is precisely the perturbative part of the expression
for the central charge as it appears in the 
$\Pi$-stability condition \cite{Douglas:2000ah} for general B-type branes
\footnote{This is true at least for space filling branes in case we consider also non-abelian fields. Of course our analysis has nothing to say about lower-dimensional non-abelian branes.}. 
To our knowledge the form of this expression
has never been derived from first principles. Rather, we understand that the 
central charge has been designed in such  a way as to keep in analogy with 
the well-known RR-charge of the B-brane as seen in the Chern-Simons action - 
it is simply assumed that in the geometric limit, the two quantities 
coincide\cite{Aspinwall:2004jr}.

We find it quite interesting though not unexpected that, starting from the well-known Green-Schwarz
anomaly terms,
our four-dimensional effective field
theory analysis leads precisely to the perturbative part
of the $\Pi$-stability condition for B-type branes.

Equation (\ref{SUSY}) is also the integrability condition for the
non-abelian generalisation of the MMMS equation for D9-branes
in a curved background. The abelian version of this equation has been 
proven (without the curvature terms) in \cite{Marino:1999af} starting 
from the DBI action of a single D-brane and it has been confirmed by a 
world-sheet calculation in \cite{Kapustin:2003se}.
Up to now it is strictly speaking only a conjecture that it can easily be 
generalised to (\ref{SUSY})\cite{Minasian:2001na,Enger:2003ue}. However, 
our analysis relies exclusively on  quantities of the four-dimensional 
${\cal N} = 1$ effective supergravity theory, the one-loop FI-term and 
the holomorphic gauge kinetic function.
 In particular, the non-renormalization theorems guarantee the absence of 
further perturbative corrections, thus dictating  (\ref{SUSY}) as the 
perturbatively exact integrability condition at least for D9-branes.
The absence of a stringy one-loop correction was shown in 
\cite{Poppitz:1998dj}. Of course, there will be additional non-perturbative
corrections, which in the $g_s \rightarrow 0$ limit make out  the complete $\Pi$-stability
expression \cite{Douglas:2000ah}. 

The integrability condition is not yet sufficient for supersymmetry
preservation, 
but has to be supplemented by the correct 
stability condition. This will be the direct generalisation 
of $\mu$-stability, which is the valid notion of stability only at  leading order
in $\alpha'$ and $g_s$. To investigate this point further, we have to know the
local supersymmetry equation for non-abelian D9-branes underlying (\ref{SUSY}). At
first sight, this seems to be beyond the scope of our supergravity analysis:  

All we can say for sure starting from (\ref{SUSY}) is that the local SUSY
condition 
for D9-branes has to be of the form
\bea
\label{SUSYloca}
\left[{\rm Im}\left( e^{i\varphi}\, e^{J \, {\rm id}  +i {\cal F}}\, \sqrt{\hat A({\cal M})}\right) \right]_{top} +  d \alpha_5 =0, \nonumber
\eea  
where $\alpha_5$ is a globally defined 5-form so that $d\alpha_5$ is gauge covariant. 
At least for compactifications on genuine Calabi-Yau manifolds, where $dJ=0$ and $dH=0$,
we cannot find 
any 5-form of this type which is also invariant under
the axionic $U(1)$ gauge symmetry $B\to B+d\chi$, $A\to A - \chi$ and
does  lead to a non-vanishing $d\alpha_5$.

Therefore, we would like to conclude that the possible correction
$d\alpha_5$ is absent and that indeed the local supersymmetry condition
is given by
\bea
\label{SUSYloc}
\left[{\rm Im}\left( e^{i\varphi}\, e^{J \, {\rm id}  +i {\cal F}}\, \sqrt{\hat A({\cal M})}\right) \right]_{top}  =0.
\eea

The notion of stability relevant for (\ref{SUSYloc})  has been 
analysed in \cite{Enger:2003ue} and been  called $\pi$-stability 
(to stress that it is only the perturbative part of $\Pi$-stability). 
In particular, the authors have shown that (\ref{SUSYloc}) has a unique solution 
precisely if the bundle is stable with respect to the deformed slope 
\bea
\label{cite}
\pi({V}) = {\rm Arg} \left( \int_{\cal M} {\rm tr}_{U(N)}
\left[ e^{J \, {\rm id}  +i {\cal F}}\,
         \sqrt{\hat A({\cal M})}\right] \right),
\eea
i.e. the phase of the central charge. For supersymmetric configurations, we 
need to ensure that all objects are BPS with respect to the same supersymmetry 
algebra. This is guaranteed by the integrability condition (\ref{SUSY}) 
(with $\varphi =0$ in our case).

We have phrased our discussion of stability on the Type I respectively Type IIB side. It is clear, though, that all concepts translate directly 
into heterotic language. In particular, it would be exciting to identify the S-dual non-perturbative effects which make out heterotic $\Pi$-stability. This would necessitate also a further study of instantons in the $E_8 \times E_8$-string, analysed perturbatively in \cite{Blumenhagen:2005ga}. 
 
To conclude this brief interlude, we would like to emphasise that in practical applications, it remains of course a 
very difficult task to explicitly 
prove stability of the bundles, as is the case already for $\mu$-stability.
Note, however, that a supersymmetric bundle still has vanishing $\pi$-slope,
so that, as with $\mu$-stability, $H^0({\cal M},V)=0$ is a 
necessary condition
for a $U(N)$ bundle to be $\pi$-stable.   
Apart from this non-trivial hint towards stability, one may  
only be able to show that the integrability condition (\ref{SUSY}) is 
satisfied.   
Taking this pragmatic point of view, just as has been customary in most explicit constructions in the literature so far, it is now possible to construct
orientifolds with magnetised D-branes for the case
that the D9-branes support not only abelian bundles
but also non-abelian ones. 
This opens up a whole plethora of new model building
possibilities and we will give two simple though not realistic examples
in section~\ref{SecEx}.

\section{Examples}
\label{SecEx}

As we have stressed several times already, model building in Type I string theory has mainly focused on 
compactifications involving abelian bundles on manifolds with almost trivial tangent bundle, i.e. the torus or 
orbifolds thereof. This, however, is clearly only the tip of the iceberg of possible constructions. Let us 
give a very simple illustration of how the use of non-abelian bundles on general Calabi-Yau manifolds admits 
new solutions to the string consistency conditions. We will formally work on the heterotic side of the story,
 but as emphasised, the picture is easily translated to Type I theory.

\subsection{A model on the Quintic}

Consider for simplicity the most basic Calabi-Yau manifold, 
the Quintic, with  Hodge numbers $(h_{21}, h_{11}) = (101, 1)$, intersection form 
\bea
I_3 = 5 \, \eta^3
\eea
and 
\bea
c_2(T) = 10\, \eta^2.
\eea
Having only one K\"ahler modulus, the Quintic nicely illustrates the virtue of non-abelian bundles. 
Namely, in order to cancel the positive contribution of the tangent bundle to the tadpole equation (\ref{TP}), 
we inevitably have to consider bundles with non-abelian structure group, since line bundles contribute in this case with the 
same sign as the tangent bundle.  
Following the well-known construction of vector bundles via exact sequences as reviewed in 
\cite{Blumenhagen:2005ga}, we therefore turn on the non-abelian bundle $V$ defined by the exact sequence
\bea 
\label{seqQ}
    0\to V   \stackrel{}{\rightarrow} 
{\cal O}(1)^{\oplus 5}\vert_{\cal M}  \to {\cal O}(3) \oplus  {\cal O}(4)\vert_{\cal M}\to 0,
\eea
from which the Chern characters are computed to be
\bea
c_1(V) = -2\, \eta, \qquad
\ch_2(V) = -10\, \eta^2, \qquad  
\ch_3(V) = -{43 \over 3}\, \eta^3.  \nonumber
\eea
The resulting structure group is $G=U(3)$ and the observable gauge group $H=SO(26) \times U(1)$
with the chiral spectrum displayed in table~\ref{TQuintic}. The abelian factor is anomalous and
merely survives as a global symmetry.

\begin{table}[htb]
\renewcommand{\arraystretch}{1.5}
\begin{center}
\begin{tabular}{|c||c|c|}
\hline
\hline
reps. & Cohomology & $\chi$\\
\hline \hline
$(\1)_2$ & $H^{\ast}({\cal M}, \bigwedge^2 V)$ & 155 \\
$({\bf 26})_1$ & $H^{\ast}({\cal M}, V)$ & -80 \\ 
\hline
\end{tabular}
\caption{\small Massless spectrum of a $H=SO(26)\times U(1)$ model on the Quintic.}
\label{TQuintic}
\end{center}
\end{table}

The one-loop corrected DUY equation~(\ref{EDUY1}) leads to the freezing
\bea
r= \sqrt{91\over 6} \, g_s,
\eea
where the K\"ahler class has been expanded as $J=\ell_s^2\, r \,\eta$.
For $g_s=0.7$ one gets $r=2.7$, so that it is possible to freeze
a combination of the string coupling and the radius such that both remain
in the perturbative regime. 

The one-loop corrected gauge couplings are given by
\bea
{1 \over g^2_{SO(26)}}\biggr\vert_{1-loop} &=& \frac{95}{18\pi} r \sim 4.5, \nonumber\\
{1 \over g^2_{U(1)}}\biggr\vert_{1-loop} &=& \frac{245}{3\pi} r \sim 70 \nonumber
\eea
for the above value of the string coupling.

\subsection{An $SO(16)$ model on a two-parameter CICY}
\label{SecEx2}

The second example involving supersymmetric non-abelian bundles is defined on the CICY
\bea
         {\cal M}=\matrix{ \IP_3 \cr  \IP_1 \cr}\hskip -0.1cm\left[{\matrix{ 4\cr 2\cr}}\right] 
\eea      
with Hodge numbers $(h_{21},h_{11})=(86,2)$, for which the intersection form
\bea
I_3=2\eta_1^3 + 4\, \eta_1^2\eta_2
\eea
and the second Chern class of the tangent bundle
\bea
c_2(T)=6\eta_1^2+8\eta_1\eta_2
\eea
were computed e.g. in~\cite{Blumenhagen:2005ga} in terms of the K\"ahler forms $\eta_1, \eta_2$ of the ambient product space.
This time, we turn on two  $U(4)$ bundles $V_i$ ($i=1,2$) described both by the exact sequence
\bea
0 \rightarrow V_i \rightarrow  {\mathcal O}(1,0)^{\oplus2}\vert_{\cal M} \oplus {\mathcal O}(0,1)^{\oplus4}\vert_{\cal M} 
\rightarrow {\mathcal O}(2,1)^{\oplus2}\vert_{\cal M}   \rightarrow 0.
\eea
The topological data specifying $V_i$ are therefore
\bea
c_1(V_i)=-2 \eta_1 + 2 \eta_2, \quad \ch_2(V_i)= -3\eta_1^2 -4 \eta_1\eta_2, \quad \ch_3(V_i)=-\frac{7}{3}\eta_1^3-3\eta_1^2\eta_2.
\eea
The sum of these two bundles satisfies the tadpole condition, and the observable non-abelian gauge group is $SO(16)$.
The linear combination $U(1)_1-U(1)_2$ is anomaly-free while its orthogonal abelian factor is anomalous.
The chiral spectrum is given in Table~\ref{TCICY2}.
\begin{table}[htb]
\renewcommand{\arraystretch}{1.5}
\begin{center}
\begin{tabular}{|c||c|c|}
\hline
\hline
reps. & Cohomology & $\chi$\\
\hline \hline
$({\bf 16})_{1,0}$ & $H^{\ast}({\cal M}, V_1)$ & -24 \\
$({\bf 16})_{0,1}$ & $H^{\ast}({\cal M}, V_2)$ & -24 \\
$({\bf 1})_{1,1}$ & $H^{\ast}({\cal M}, V_1\otimes V_2)$ & -120 \\
$({\bf 1})_{1,-1}$ & $H^{\ast}({\cal M}, V_1 \otimes V_2^*)$ & 0 \\
\hline
\end{tabular}
\caption{\small Massless spectrum of a $H=SO(16)\times U(1) \times U(1)_{massive}$ model on a CICY with $h_{11}=2$.}
\label{TCICY2}
\end{center}
\end{table}

The supersymmetry conditions for both bundles are of course identical and lead to
\bea
\label{Florian}
r_1 \left(4 r_2 - r_1\right) = \frac{67}{6} g_s^2
\eea
 in terms of our usual expansion of the K\"ahler from as $J=\ell_s^2 \sum_{i=1}^2 r_i \eta_i$.
The gauge couplings are given by
\bea
{1 \over g^2_{SO(16)}}\biggr\vert_{1-loop} &=& \frac{1}{\pi}\frac{\frac{50}{18}r_1^2+8r_1r_2-2r_2^2}{4r_2-r_1} , \nonumber\\
{1 \over g^2_{U(1)_i}}\biggr\vert_{1-loop} &=& \frac{1}{\pi}
\frac{\frac{2}{9}r_1^2 +140 r_2^2 +32 r_2^2}{4r_2-r_1}, \quad i=1,2
\eea
where we have used (\ref{Florian}) to eliminate the string coupling.

We have demonstrated the possibility of reducing the rank of the observable gauge group, in particular of the unwanted $SO(2M)$ remnant, with the help of non-abelian bundles. Of course more realistic models would have to involve in addition line bundles which would give rise to $SU(N_i)$ factors as discussed. The search for appealing models of this kind is beyond the scope of this publication and remains for future investigation.

\vbox{
\section{Phenomenological aspects}
\label{SecPhen}

In this section we briefly address  some of the 
qualitative phenomenological
aspects of these heterotic $SO(32)$ compactifications with non-abelian and abelian 
bundles. Some of them have already been mentioned in the sections before, but
we nevertheless list them here for completeness. When choosing just abelian bundles, 
the phenomenological features are very similar to the intersecting D-brane
models in Type I string theories. Please keep in mind that
the methods developed so far are valid in the small coupling, large radius
regime and that away from the perturbative limit certain low-energy
parameters also receive world-sheet and stringy instanton corrections.\\
}

\noindent 
{\it The size of the string scale}

The phenomenological consequences of these heterotic string models
drastically depend on the value of the string scale. 
It is known that in Type I models with non-spacetime filling
D-branes the string scale can in principle be anywhere between
the TeV range and the Planck scale. 
For the heterotic string the tree-level gravitational and gauge couplings
are of the order
\bea
\label{scales}
       M^2_{pl}\sim {M_s^8\, V_6 \over g_s^2},\quad\quad\quad\quad
  {1\over g_{YM}^2}\sim {M_s^6\, V_6\over g_s^2},
\eea
which immediately implies $M_{pl}^2 \sim M_s^2/g^2_{YM}$. For gauge couplings of order one
at the unification scale, the string scale is therefore of the order of the Planck scale.

Let us now analyse the possible consequences of the higher string loop corrections. 
It is known that the Einstein-Hilbert term does not
receive any stringy loop corrections \cite{Antoniadis:1992sa}, 
whereas for the 
gauge coupling the exact  one-loop corrected expression is of the form
\bea
\label{loopscales}
   {1\over g_{YM}^2}\sim {M_s^6\, V_6\over g_s^2} - M_s^2\, V_2\, \beta.
\eea
Here $V_2$ denotes the two-cycle volume (transversal to the four-form flux) and
$\beta$ can be expressed by the various second Chern-characters appearing
in (\ref{Gauged}) and is essentially an integer of order $10-10^2$ for
most known examples. Note for more than one K\"ahler modulus, the one-loop
correction actually contains a sum over all two-cycle volumes.
For $\beta>0$ (as is the case for the $SO(16)$ gauge coupling in example~\ref{SecEx2})
and assuming the gauge coupling to be of order one, we
get
\bea
\label{plms}
          {M^2_{pl}-M^2_s\over M_s^2}\sim  M^2_s\, V_2\, \beta.
\eea
This tells us that $M_s$ can be significantly smaller than the
Planck scale  as long as the 2-cycle volume $V_2$ is large in units of $M_s$.
To be more precise, we have to make sure that the lightest Kaluza-Klein
mode of the Standard Model gauge sector is not lighter than the weak
scale $M_w$. Thus, identifying $V_2=1/M_w^2$ we can lower the string scale
down to
\bea
      M_s\sim {\sqrt{M_{pl}\, M_w}\over \beta^{1\over 4}} \simeq {10^{11}\over
        {\beta^{1\over 4}}}
      {\rm GeV}.
\eea  
The first equation in (\ref{scales}) implies that the ``longitudinal''
volume $V_4$, defined as $V_6=V_2\, V_4$, is of order $\beta$
in string units and therefore large enough $\beta$ still
allows $V_4$ in the perturbative regime. For $\beta<0$ both the
tree-level and the one-loop correction in (\ref{loopscales}) are positive and we arrive
at the usual  conclusion that $M_s\simeq M_{pl}$. 

To summarize, provided that in a heterotic string compactification
we can engineer the Standard Model gauge group via bundles
with positive combinations of the second Chern classes $\beta$,
the one-loop corrections allow us to lower the string scale
down to the intermediate regime if the background
is  non-isotropic and has large (transversal) two-cycles. 
Whether this scenario can actually be realized in Standard-like models remains to be
seen but it again shows that we ought to carefully reevaluate our 
prejudice on heterotic string compactifications. \\

\noindent 
{\it The observable gauge group}

Clearly, the observable gauge group $H$  is the commutant of the structure
group $G$
in $SO(32)$. For our  first choice of the structure group this gauge group was
\mbox{$H=SO(2M) \times \prod_{j=1}^K SU(N_j)$} extended by the 
anomaly free part of the $U(1)^{K+L}$ gauge factors. Therefore the effect of truly
non-abelian $U(N_m)$ factors in the structure group is to reduce the rank of the
initial gauge group $SO(32)$.

In general there are multiple anomalous $U(1)$ gauge factors, which become
massive via some bilinear couplings to the K\"ahler-axions and the
universal axion. The scale of each mass term is of the order of
the string scale; however for multiple anomalous $U(1)$s we can arrange for mass eigenvalues in the entire phenomenologically acceptable 
range from the weak scale up to the string scale. 
Therefore, one might consider the existence of additional massive $U(1)$
gauge factors as the main model independent prediction of this
kind of string compactifications. For Standard-like models one has to ensure
that there exists a $U(1)_Y$ which remains massless after the 
Green-Schwarz mechanism has been taken into account. 
Note that for the $E_8\times E_8$ heterotic string there exist choices
of the structure group which do not lead to any observable $U(1)$.\\

\noindent
{\it The  matter spectrum}

The chiral matter comes in various bifundamental and (anti-)symmetric
representations of the observable gauge group, and the net number
of generations is given by the Euler characteristic of the tensor product
of the bundles in question (see Table 1). For the general embedding we
are considering there never occurs any spinor representation of
the $SO(2M)$ part of the observable gauge group. 
Thus, for realising $SO(10)$ GUTs one has to consider
compactificatons of the $E_8\times E_8$ heterotic string.

If one wishes to determine the complete chiral
and non-chiral spectrum,  knowledge of the index is of course not enough. One
really has to compute the various cohomology classes. Here also additional
massless states can appear in the singlet and adjoint representations of the
gauge group. These are additional potential moduli fields and correspond in the
S-dual Type I formulation to open string moduli.
By working at a generic point in moduli space we expect that the amount of non-chiral 
matter is reduced  \cite{Braun:2005bw,Braun:2005ux} as compared to  the mostly studied 
special points in moduli space
like orbifolds, free-fermion constructions and Gepner models.\\

\vbox{
\noindent
{\it The  gauge couplings}

We have derived the perturbative expressions for the non-abelian and 
abelian gauge couplings. Besides the - up to normalisation - universal tree-level result there appears
a one-loop threshold correction.
Including these corrections, both the non-abelian and the abelian
gauge couplings depend on the K\"ahler moduli and are non-universal, which is also expected from
the S-dual Type I picture. Quite analogously also, the gauge kinetic functions
of the anomaly-free abelian factors are given by the corresponding linear combinations 
of the gauge kinetic functions of the original $U(1)$ factors.
For concrete Standard-like models taking into account the light
charged matter fields, one has to check case by case whether
these moduli dependent gauge couplings can give rise  to gauge coupling unification
at the string scale. \\
}

\noindent
{\it Supersymmetry}

For a specific observable gauge group and charged matter content, i.e.
without giving VEVs to charged scalars, the D-terms are determined
by the Fayet-Iliopoulos terms. As we have shown these contain
a tree-level and a stringy one-loop correction  and do depend
on the K\"ahler moduli. Therefore, for $U(N)$ bundles supersymmetry implies that 
certain combinations of the dilaton and the K\"ahler moduli
get fixed. For pure $SU(N)$ bundles with non-vanishing third
Chern class, however, the tree level FI-term vanishes and
one gets only a moduli independent one-loop induced FI-term. 
It follows that in this case supersymmetric vacua can at best
be reached by turning on VEVs for charged scalars. 

We have no real control over the F-terms, where
world-sheet instantons can generate   non-vanishing contributions.
It is known that for $(0,2)$ models which admit an anomaly-free
linear sigma model description no world-sheet instantons
destabilize the background \cite{Silverstein:1995re,Basu:2003bq,Beasley:2003fx}. 
However, this is not necessarily true
for the general tadpole cancelling configurations of abelian and non-abelian
gauge bundles we consider here, where the left-moving $U(1)$ on the
world-sheet is anomalous. It would be interesting to systematically study
these globally anomalous linear sigma models. \\

\noindent
{\it Fluxes and moduli stabilisation}

For the heterotic string one can also turn on a background $H_3$ form
flux. This induces a superpotential of the form \cite{Becker:2003yv,Becker:2003gq,Cardoso:2003af}
\bea
W_{flux} \sim \int \Omega \wedge (H + i \; dJ) 
\eea\
and allows to stabilise the complex structure moduli. Here we have also
included the torsion piece, which of course vanishes in the
supergravity limit. 
In contrast to the Type IIB case, the dilaton does not appear
in the superpotential and therefore cannot be stabilised in this way.
As a result, stabilisation of  the K\"ahler, dilaton and bundle
moduli seems only  possible by taking into account 
non-perturbative contributions to the superpotential.~\cite{Curio:2001qi,Becker:2004gw,Curio:2005ew}\\

\noindent
{\it Yukawa couplings}

For phenomenological purposes it is very important to compute the Yukawa couplings for this class
of models. The physical Yukawa couplings can only be read off if the
kinetic terms are canonically normalised, which means
that the physical Yukawa couplings involve both the
K\"ahler potential and the superpotential. The K\"ahler potential
 for charged fields might be  hard to derive, but the superpotential
should already allow us to find some selection rules for the
non-vanishing couplings. At large radius these selection rules are expected
to be given
by the cohomology ring $H^q({\cal M},{\cal W})$, i.e.  
\bea
H^p({\cal M},{\cal W}_1) \times H^q({\cal M},{\cal W}_2)  \times H^r({\cal M},{\cal W}_3) \rightarrow H^3({\cal M},{\cal O}) =\mathbb{C},
\nonumber 
\eea
where the  bundles ${\cal W}_i$ are to be taken  from Table~\ref{Tchiral1}.
For anomaly-free linear sigma models these cohomology rings could 
be computed as  chiral (topological)  rings of the underlying conformal
field theory \cite{Blumenhagen:1995ew,Adams:2005tc,Katz:2004nn}. 
It remains to be investigated how these couplings
can be explicitly found for more general bundles corresponding to
anomalous linear sigma models. \\

\noindent
{\it Soft supersymmetry breaking terms}

Soft supersymmetry breaking in the observable gauge sector can either be achieved
by some breaking in the hidden sector of a given model, e.g. via gaugino condensation, 
or through suitable three-form fluxes.
Once one knows the gauge kinetic function, the K\"ahler potential and
the superpotential, the induced soft-terms can be computed by
the general supergravity formulas \cite{Ibanez:1992hc,Kaplunovsky:1993rd,Brignole:1993dj}, 
which are parameterised
by the supersymmetry breaking auxiliary fields $F_s$  in the
chiral superfields. 
\\

\noindent
{\it Stability of the Proton}

As in the intersecting D-brane scenario, for a concrete realization of the
Standard Model baryon and lepton number might appear as gauged anomalous 
$U(1)$ symmetries \cite{Ibanez:2001nd} which
become massive via the Green-Schwarz mechanism and therefore survive
as perturbative global symmetries. 
Clearly in such models there would be no problem with rapid proton decay.

\section{Conclusions}
\label{SecCon}

The aim of this article is to generalise the construction
of string compactifications on smooth  Calabi-Yau spaces
with non-abelian and in particular abelian vector
bundles to the case of the $SO(32)$ heterotic string.
We worked out the generalised Green-Schwarz mechanism
in very much detail and computed the one-loop
threshold corrected gauge kinetic functions and
the FI-terms in the perturbative regime. 
In this sort of topological sector, 
we found completely S-dual features
on the heterotic and Type I sides. The only difference,
dictated by S-duality, is that certain 
$\alpha'$ corrections in Type I theory are mapped
to string loop corrections on the heterotic side.

We conclude that for smooth heterotic/Type I compactifications
all features are in accordance with S-duality and
that the seeming differences can be traced back to
the restricted set of examples one has 
considered so far and probably to some subtleties related
to the fixed points of orbifold models. 
Having discussed the smooth case in detail, it still remains to 
be understood 
how heterotic orbifolds and free-fermion constructions
precisely fit into our approach.

As a byproduct of our effective four dimensional
analysis, we were able to confirm from
first principles, i.e. from the celebrated
Green-Schwarz terms,  the perturbative part of the 
$\Pi$-stability condition for B-type branes. 

So far we have only laid the foundation and qualitatively 
discussed a couple of phenomenological aspects of this 
class of $SO(32)$ heterotic compactifications.  
A  more
systematic study of all the string vacua with abelian
and non-abelian bundles has to follow. More examples and the question
of whether Standard-like models can be constructed this
way remain for future investigation. 
This might
include a  study of the statistics of the vacua of this type. 
It would also be  interesting to investigate for concrete models
the possibility of moduli stabilisation via fluxes
\cite{Becker:2002sx,Cardoso:2002hd}.

 \vskip 1cm
 {\noindent  {\Large \bf Acknowledgements}}
 \vskip 0.5cm 
It is a great pleasure to thank Mirjam Cveti\v{c}, Michael Douglas, Florian Gmeiner, 
 Mariana Gra\~{n}a, Claus Jeschek,  Dieter L\"ust and Stephan Stieberger for helpful discussions.
 \vskip 2cm

\appendix

\section{Anomalies in terms of Euler characteristics}
\label{AppEulerExp}

The cubic non-abelian, mixed and pure abelian anomalies can be expressed in terms of the Euler characteristics
as follows for the case that the structure group is $G=\prod_{m=K+1}^{K+L} SU(N_m)\times U(1)^{K+L}$,
\bea
{\cal A}_{SU(N_i)^3} &\sim& A_{SU(N_i)^3}=
(N_i-4)\chi(L_i^2) + \sum_{j \neq i} N_j \left(\chi(L_i \otimes L_j) + \chi(L_i \otimes L_j^{-1})   \right)  
\nonumber \\
&&+\sum_m \left(\chi(V_m \otimes L_i) +\chi(V_m^{\ast} \otimes L_i)      \right)
+ 2M \chi (L_i), \nonumber \\
{\cal A}_{U(1)_i-SU(N_i)^2} &\sim& N_i \chi(L_i^2)+ A_{SU(N_i)^3}, 
\nonumber \\
{\cal A}_{U(1)_i-SU(N_{j\neq i})^2} &\sim& A_{U(1)_i-SU(N_{j\neq i})^2} 
= N_i \left(\chi(L_i \otimes L_j) + \chi(L_i \otimes L_j^{-1})   \right),  
\nonumber \\
{\cal A}_{U(1)_m-SU(N_i)^2} &\sim&  A_{U(1)_m-SU(N_i)^2} 
= \chi(V_m \otimes L_i) -\chi(V_m^{\ast} \otimes L_i),  
\nonumber \\
{\cal A}_{U(1)_i-G^2_{\mu\nu}} &\sim& 3 N_i  \chi(L_i^2) + A_{SU(N_i)^3} , 
\nonumber \\ 
{\cal A}_{U(1)_m-G^2_{\mu\nu}} &\sim& A_{U(1)_m-G^2_{\mu\nu}} 
 = 2 \chi(\mbox{$\bigwedge^2 V_m$} )
+\sum_i N_i \left(\chi(V_m \otimes L_i) -\chi(V_m^{\ast} \otimes L_i)      \right)
 \nonumber \\
&&+\sum_{n\neq m} \left(\chi(V_m \otimes V_n ) +\chi(V_m \otimes V_n^{\ast} )   \right)
+2M \chi (V_m ), \nonumber \\
{\cal A}_{U(1)_i^3} &\sim&  3 N_i^2 \chi(L_i^2)+ N_i  A_{SU(N_i)^3} ,  
\nonumber \\
{\cal A}_{U(1)_i-U(1)_{j\neq i}^2} &\sim&  N_jA_{U(1)_i-SU(N_{j\neq i})^2} ,  
\nonumber \\
{\cal A}_{U(1)_i-U(1)_m^2} &\sim&  N_i\left(\chi(V_m\otimes L_i) +\chi(V_m^{\ast} \otimes L_i) \right),  
\nonumber \\
{\cal A}_{U(1)_m-U(1)_i^2} &\sim&  N_i A_{U(1)_m-SU(N_i)^2},
 \nonumber \\
{\cal A}_{U(1)_m^3}  &\sim&  6 \chi(\mbox{$\bigwedge^2 V_m$} ) + A_{U(1)_m-G^2_{\mu\nu}} , 
\nonumber \\
{\cal A}_{U(1)_m-U(1)_n^2} &\sim& \chi(V_m \otimes V_n ) +\chi(V_m \otimes V_n^{\ast}), \nonumber \\
{\cal A}_{U(1)_i-SO^2} &\sim& N_i \chi(L_i), \nonumber \\ 
{\cal A}_{U(1)_m-SO^2} &\sim&  \chi (V_m). \nonumber 
\eea
If  $SO(2M)$ also belongs to the bundle, the following modifications occur:
in $A_{SU(N_i)^3}$ we make the  replacement
\mbox{$2M \chi (L_i) \rightarrow \chi(V_0 \otimes L_i)$}; in
$A_{U(1)_m-G^2_{\mu\nu}} $ we substitute  
\mbox{$2M \chi (V_m ) \rightarrow\chi (V_0 \otimes V_m )$}.

\section{Collection of trace identities}
\label{AppTrace}

The detailed computation of the GS counter terms depends on the concrete embedding of the structure group into $SO(32)$. 
In this appendix we list some useful trace identities which enter the analysis of the GS mechanism in section~\ref{GSM}.
We restrict ourselves here to the choice of bundles made in (\ref{decomp1}), 
but completely analogous expressions arise for all other embeddings.

To shorten the notation, we treat the internal background $U(N_m)$-bundles $V_m$ and the line bundles $L_i$ 
on the same footing and denote them simply by $V_x$, $x \in \{1, \ldots, K+L\}$.
It turns out to be  convenient to decompose the corresponding field strength  ${\ov{F}}_x$ 
into its non-abelian $SU(N_x)$-part $\widehat{\ov F}_x$ (where present) and the diagonal $U(1)$-part ${\ov f}_x$ as
\bea
\label{F}
{\ov { F}}_x =\widehat{\ov  F}_x + {\ov f}_{x}\,{\rm I}_{N_x \times N_x},
\eea
with ${\rm I}_{N_x \times N_x}$ denoting the identity matrix.
In other words, it is understood that the non-abelian part of ${\ov { F}}_x$ is simply absent for $x \in \{1, \ldots, K\}$, where the background bundles are merely line bundles. However, even though in this case the $SU(N_x)$ is part of the external gauge theory, we insist on the presence of the identity matrix as a remnant to compactly keep track of various factors of $N_x$.

Let us note explicitly that (\ref{F}) implies the obvious identities
\bea
 {\rm tr} {\ov {F}}_{x} &=& N_x {\ov f_x}, \nonumber \\
 {\rm tr} {\ov {F}}^2_{x} &=& {\rm tr} \widehat{\ov F}^2_{x}+ N ({\ov f_x})^2, \nonumber \\
{\rm tr} {\ov {F}}^3_{x} &=& {\rm tr} \widehat{\ov F}^3_{x}+ 3 {\ov f_x} \, {\rm tr}\widehat{ \ov F}^2_{x} + N_x ({\ov f_x})^3,
\eea
where all traces are in the fundamental representation.

Analogously, unbarred quantities denote the respective four-dimensional gauge field strengths.

With these conventions, it is a simple exercise to decompose the traces in the adjoint of $SO(32)$, Tr, of the various combinations of external and internal fields appearing in the GS counter terms into traces over the fundamental of the $U(N_x)$-factors,

\bea
 {\rm Tr} F{\ov F}^3 &=& 12\sum_{x=1}^{K+L} f_x \wedge \left( 4 \; {\rm tr} {\ov {F}}_x^3 
+ {\rm tr} {\ov {F}}_x   \sum_{y=1}^{K+L} {\rm tr} {\ov {F}}_y^2\right), \nonumber\\
 {\rm Tr} F^2 {\ov F}^2 &=&
4 \sum_{i=1}^{K} {\rm tr}_i \widehat{F}^2 \wedge \Bigl( 12\; (\ov f^i)^2 + \sum_{x=1}^{K+L} {\rm tr} {\ov {F}}_x ^2 \Bigr) \nonumber \\
&+& 4 \sum_{x=1}^{K+L} \; (f_x)^2 \wedge \left(12\;  {\rm tr} {\ov {F}}_x^2 
+ N_x   \sum_{y=1}^{K+L}  {\rm tr} {\ov{F}}_y^2\right) \nonumber \\
&+&8\sum_{x,y=1}^{K+L} \, f_x \,  f_y \wedge  {\rm tr}{\ov {F}}_x  \,  {\rm tr}{\ov {F}}_y
+\; 2 \; {\rm tr}_{SO(2M)} F^2 \wedge \sum_{x=1}^{K+L} {\rm tr} {\ov {F}}_x^2   , \nonumber\\
{\rm Tr} F^2&=& 30 \, {\rm tr}_{SO(2M)} F^2 + 60 \sum_{x=1}^{K+L}   {\rm tr}{F}_x^2     , \nonumber\\
{\rm Tr}F {\ov F} &=& 60 \sum_{x=1}^{K+L}  f_x\wedge {\rm tr}{ \ov{F}}_x,  \nonumber\\ 
{\rm Tr} {\ov F}^2 &=& 60 \sum_{x=1}^{K+L} {\rm tr} {\ov {F}}_x^2  .
\eea

It follows immediately that the GS counter terms computed in section~\ref{GSM} have precisely the right form to cancel the field theoretic anomalies.

\clearpage
\nocite{*}
\bibliography{SOrev}
\bibliographystyle{utphys}

\end{document}